# Multiple interfaces in diffusional phase transitions in binary mesogen-non-mesogen mixtures undergoing metastable phase separations

Ezequiel R. Soulé[1,2], Cyrille Lavigne[2,3], Linda Reven[3] and Alejandro D. Rey[2*]

1. *Institute of Materials Science and Technology (INTEMA), University of Mar del Plata and National Research Council (CONICET), J. B. Justo 4302, 7600 Mar del Plata, Argentina*

2. *Department of Chemical Engineering, McGill University, Montreal, Quebec H3A 2B2, Canada*

3. *Center for Self-Assembled Chemical Structure (CSACS), Chemistry Department, McGill University, 801 Sherbrooke St. West, Montreal, Quebec H3A 2K6, Canada*

* corresponding author

**Abstract**

Theory and simulations of simultaneous chemical demixing and phase ordering are performed for a mixed order parameter system with an isotropic-isotropic (I-I) phase separation that is metastable with respect to an isotropic-nematic (I-N) phase ordering transition. Under certain conditions, the disordered phase transforms into an ordered phase via the motion of a double front containing a metastable phase produced by I-I demixing, a thermodynamically driven mechanism not previously reported. Different kinetic regimes are found depending on the location of the initial conditions in the thermodynamic phase diagram and the ratio between diffusional and nematic phase ordering mobilities. For a diffusional process, depending if the temperature is above or below the critical co-dissolution point, an inflection point or a phase separation takes place in the depletion layer. This phase separation leads to the formation of a second interface where the separation of the two metastable isotropic phases grows monotonically with time. The observed deviations from the typical Fickian concentration profiles are associated with strong positive deviations of the mixture from ideality due to couplings between concentration and nematic ordering. Although systems of interest include liquid crystalline nanocomposites, this novel mechanism may apply to any mixture that can undergo an order-disorder transition and demix.

## 1. Introduction

The study of phase transformations and their dynamics is an important topic in non-equilibrium thermodynamics and materials science [1-3]. In multi-component mixtures, phase transformations usually involve mass diffusion, which can be coupled to a phase ordering process if long-range molecular order is associated to one or both phases (e.g. solidification, colloidal crystallization or mesophase formation). A diffusional process can be described by the Cahn-Hilliard equation, and the phase transformation can proceed through two main mechanisms: spinodal decomposition (if the system is locally unstable to composition fluctuations), or nucleation and growth (if the system is stable against infinitesimal fluctuations). The kinetics of both mechanisms has been widely studied and the dynamic laws describing them are well known [1-3]. In the case of nucleation and growth, the process is usually described by Fick's theory, which can be considered as a sharp-interface equivalent to the continuous Cahn-Hilliard equation. In the most simple case of constant diffusivity, this model predicts the formation of a "depletion layer" in the vicinity of the interface of the growing phase. Here the concentration varies continuously with a smoothly decreasing curvature from the concentration at the interface (given by the thermodynamic equilibrium phase diagram) to the bulk concentration (given by the initial conditions), and the velocity of the interface decreases with time as $t^{-1/2}$. This model is widely used to describe many processes of industrial relevance such as alloy solidification, separation processes, dissolution or precipitation, etc [4].

When mass diffusion is coupled to phase ordering, the phase transformation process is more complex. At least two independent thermodynamic variables are necessary to fully describe each phase: concentration and order parameter (or phase-field) [3]. Concentration is a "conserved" variable (it obeys the mass conservation law), while the phase-field is non-conserved. Since the interface is not at equilibrium, the concentrations at the interface are not given by the phase diagram, and the process can show complex dynamics. The situation can be further complicated when metastable states are possible. In this case the phase transformation can proceed in stages, where metastable phases are formed before the stable ones. This is generally known as the Ostwald step rule [5, 6]. A well-known example is the case where liquid-liquid (L-L) phase separation is metastable against crystallization and the mixture first phase-separates into the two liquids and then crystallizes. This behavior has been observed in a wide range of material systems including metallic alloys [7], polymer solutions, blends and composites [6, 8, 9], colloidal dispersions[10, 11], biological systems [12], and liquid-crystalline mixtures [13-15].

Several mechanisms are available for the appearance of a metastable phase, depending on kinetic and thermodynamic factors: spinodal decomposition (in the case of a L-L phase separation), nucleation of the metastable phase, or spontaneous formation of a metastable “corona” surrounding a stable nucleus during phase growth. This work is concerned with the third process: the spontaneous formation of a metastable “corona” surrounding a stable nucleus during phase growth. The role of metastable states in the phase growth process has been theoretically studied in the context of phase-field models for the case of non-conserved order parameters (NCOP) [16-18], for conserved order parameters (COP) [19, 20], and for mixed order parameters [21, 22]. A characteristic phenomenon is the possibility of “front splitting”, which means that the original interface can split into two interfaces, generating the metastable phase. In two or three dimensions, this produces a core-shell structure where the core is the stable phase, surrounded by a corona of the metastable one. This front splitting can be spontaneous or not, and the split state can display three different evolutions: (1) a finite lifetime, (2) a quasi-stationary state or (3) an infinite lifetime, depending on the nature of the order parameters.

Previous work has described the following “kinetic front splitting” mechanism [16-22]: Due to the existence of an intermediate metastable state, the original interface, separating the stable phase, $S_1$, and the supercooled (or overheated) phase, $S_2$, can be considered to be composed by two interfaces: one separating $S_1$ and M (where M is a metastable phase), and the other separating M and $S_2$. A necessary condition for the front splitting is that the M-$S_2$ interface is faster than the $S_1$-M interface. For a single NCOP and the mixed case this is also a sufficient condition [16, 21, 22], while for a pure COP [19, 20] and some cases of multiple NCOPs [17], a finite perturbation is needed.

The process of kinetic front splitting during material transformations is schematically illustrated in Fig. 1, for the simplest possible case of a material whose thermodynamics is described by one non-conserved order parameter $\psi$, such as a nematic liquid crystal, a colloidal crystal or a crystalline solid. Figure 1(a) shows the free energy $f_b$ as a function of the order parameter $\psi$, presenting three minima: (1) $\psi_{S1}$, corresponding to the supercooled phase, (2) $\psi_{S2}$, the absolute minimum corresponding to the stable phase, and (3) $\psi_M$, corresponding to the metastable phase. Fig 1(b) shows order parameter profiles at $t = 0$ and $t > 0$, illustrating the splitting process. The transformation from $S_1$ to $S_2$ occurs by the propagation of a double front that encloses the metastable state M. The case of a conserved order parameter is similar, except that equilibrium phases are not determined by local minima, but by a double-tangent construction [19, 20]. Some of the key issues in front splitting during material transformations include: (i) the thermodynamic origin of metastability, (ii) the role of metastability in the transformation kinetics and structure, (iii) the life span of the metastable state, and (iv) the spatial extension of the metastable region.

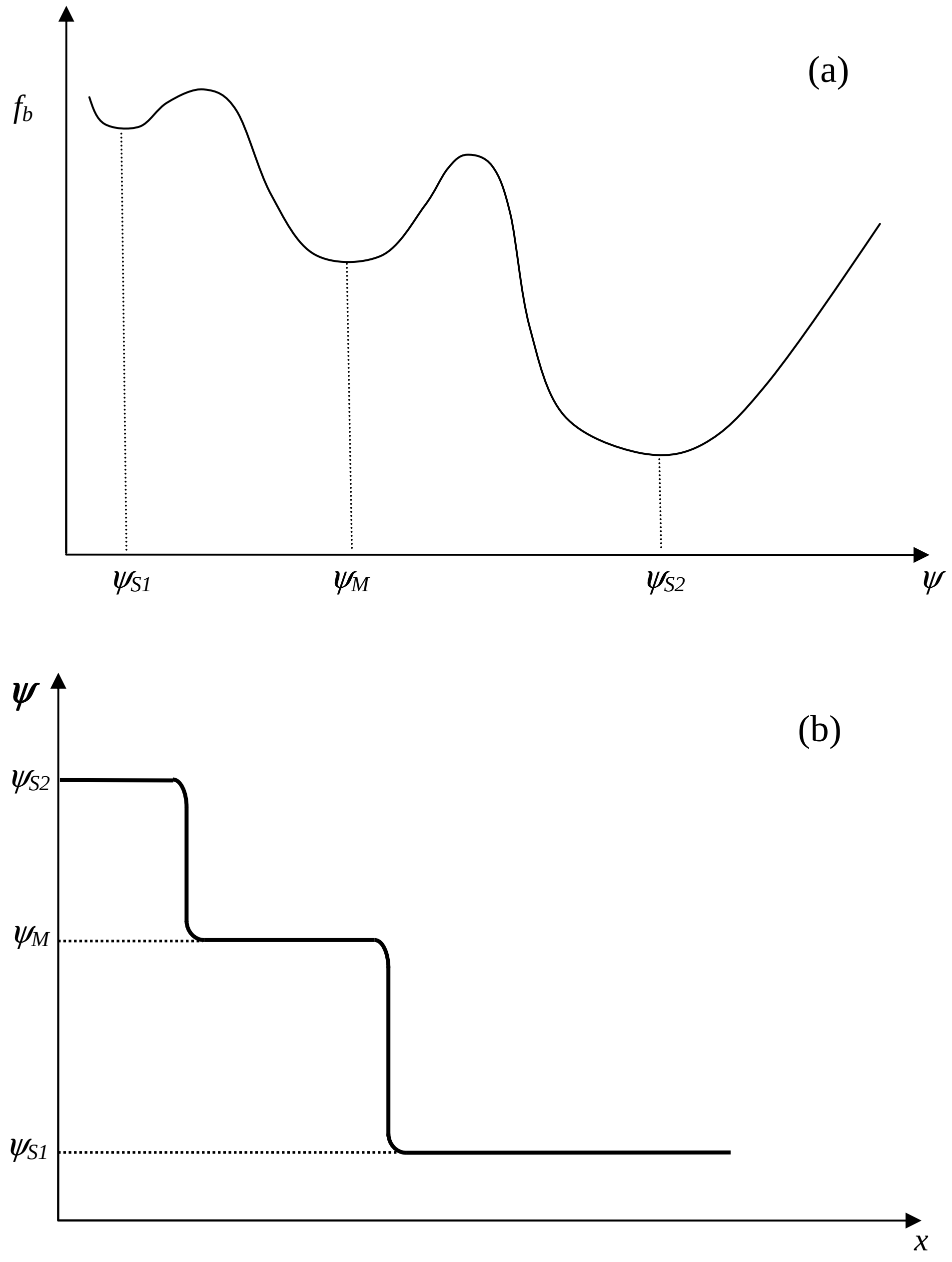

**Figure 1.** Schematic illustration of front splitting during a material transformation for the case of one non-conserved order parameter $\psi$. (a) The free energy density $f_b$, shows an absolute minimum for the stable phase, $S_2$, and a local minimum for the metastable phase M. (b) Spatial profiles of the order parameter $\psi(x)$, showing a split interface (two consecutive jumps) between $S_1$ and $S_2$, where the metastable phase M is generated between $S_1$ and $S_2$.

Here we report a novel thermodynamic mechanism that drives the appearance of multiple fronts and metastable states in systems with one conserved and one non-conserved (nematic orientational order) parameter, that were not described in any of the previous studies [16-22]. Although we use a model for nematic ordering in a mixture of spherical nanoparticles and a nematic mesogen, the results of this work are, in principle, representative of any binary mixture that can undergo an order-disorder transition and demix, so we do not focus or emphasize a specific experimental system. Instead, we highlight the generic thermodynamic and kinetic aspects of a novel process that emerges under the intersection of metastable demixing and phase ordering. It should be pointed out that the formation of double fronts is, in principle, possible in all the material systems mentioned earlier (and more), but in practice there is a competition between this mechanism and others (spinodal decomposition, faster nucleation of metastable phase), so it might be difficult to observe experimentally in quenched systems. In addition, the presence of multiple domains of complex shapes and different sizes might difficult to do a “clean” observation of a double interface as the depletion layers of different domains are irregular in shape and they overlap. This mechanism should be more relevant in directional growth experiments, where a single, flat interface is formed and kept stationary by “pulling” the fluid. In this set-up a front-splitting should be seen when the location of processing conditions in the phase diagram is adequate. Nevertheless, even if it not observed explicitly, it can profoundly affect the dynamics of phase transitions and the structure of the material (as the interfacial tension depends on the concentration difference between both phases, the presence of a metastable phase with different concentration will affect the interface tension and thus can affect the domain shape, defect structure, etc).

By performing one-dimensional transient simulations of the growth of a nematic phase in a mixture with a metastable liquid-liquid phase equilibrium, it is shown that under certain thermodynamic conditions a phase separation takes place within the depletion layer, producing the double front. This is shown schematically in Fig. 2, where we summarize the key focus and objective of this paper. The top row shows how quenching a binary mesogen (rods) /non-mesogen (circles) mixture produces a phase ordering transition. The bottom (A) schematic shows the usual transformation of the disordered phase into the ordered phase by the motion (right pointing arrow) of a single front. The bottom (B) schematic illustrates the key focus of this paper: *to characterize the transformation of the disordered phase into the ordered phase by the motion (right pointing arrow) of a double front that contains a metastable phase produced by a I-I demixing.*

The organization of this paper is as follows. Section 2 discusses the phase behavior, describing the stable and metastable phases for an example system consisting of a mixture of nanoparticles and a liquid crystal (mesogen) that presents a nematic (N) phase, with a buried isotropic-isotropic (I-I) equilibrium. Section 3 briefly describes the continuous- and sharp-interface model. Section 4 presents the simulation results for the case of a mixed dynamics, describing the new thermodynamic mechanism for the appearance of double fronts: a phase separation within the depletion layer. Section 5 describes the case of a diffusional process where section 5.1 presents simulation results, and section 5.2 presents an analytical theory based in Fick's diffusion law and compares both approaches. Finally, section 6 gives the conclusions.

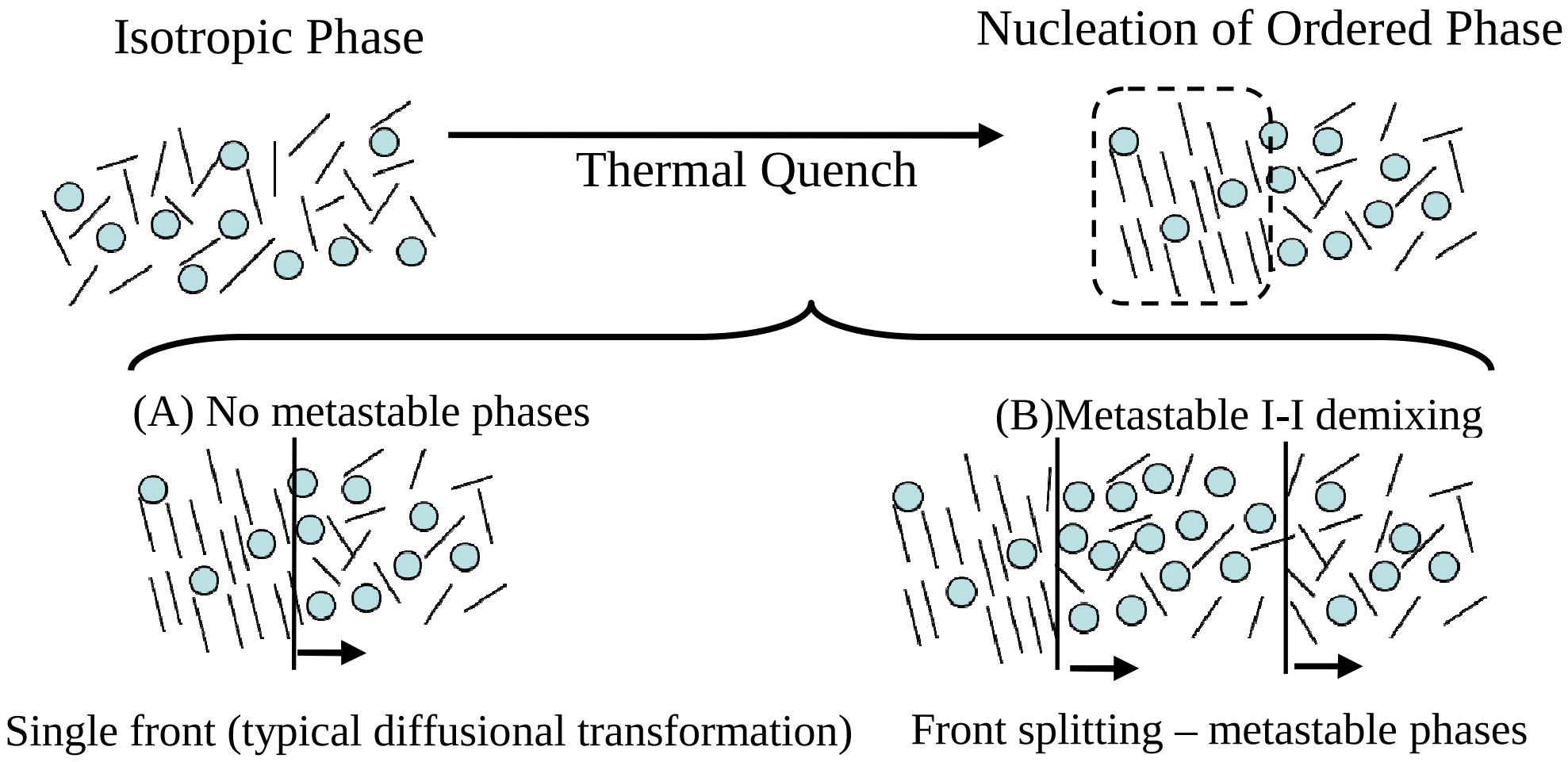

**Figure 2.** Schematic summarizing the focus and objective of this paper. The top row shows quenching a binary mesogen (rods) /non-mesogen (circles) mixture that produces a phase ordering transition. The bottom (A) schematic shows the usual transformation of the disordered phase into the ordered phase by the motion (right pointing arrow) of a single front, which is a typical diffusional transformation. The bottom (B) schematic illustrates the main objective of this paper that is to characterize the transformation of the disordered phase into an ordered phase by the motion of a double front that contains a metastable phase produced by I-I demixing, where the two isotropic phases have different concentrations.

## 2. Phase equilibrium, metastable states and phase transitions

We consider a binary system with one conserved and one non-conserved order parameter. In this work we use a specific model representing a mixture of non-self-assembling spherical particles and a calamitic thermotropic nematic liquid crystal, but as mentioned above the generic features of the results, such as interfacial structure and kinematics, are representative of any a mixture that can undergo an order-disorder transition and demix. As we will show, the phenomenon under study does not depend on the specific model used for the mixing free energy; it depends on thermodynamic conditions (metastable phases, positive deviations from ideality) that can be met in any type of material system.

The free energy of the mixture is given by a combination of Flory-Huggins theory and Carnahan-Starling equation of state for the mixing free energy, combined with Maier-Saupe theory for nematic order. The Carnahan-Starling equation of state accounts for excluded-volume interactions between particles. Molecular interactions are considered to be proportional to contact areas and they include isotropic and nematic terms (the nematic binary interactions account for anchoring at the particle surface and nano-scale distortions of the director in the vicinity of the particle). A detailed description of the model can be found in previous works [23, 24] and to avoid lengthy repetitions we quote the most relevant expressions. We emphasize that the most relevant fact is that the free energy can have two minima in the disordered state, and an additional minimum in the ordered state.

The dimensionless bulk free energy density for a nanoparticle-mesogen binary system is [23, 24]:

$$\frac{v_{ref} f_b}{RT} = \frac{\phi_P}{v_{NP}} \ln(\phi_P) + \frac{\phi_{LC}}{v_{LC}} \ln(\phi_{LC}) + \frac{\phi_P}{v_P} \frac{\left(4\phi_P - 3\phi_P^{\,2}\right)}{\left(1-\phi_P\right)^2} + \left(\chi + wS^2\right) a_p \phi_P \varphi_{LC} + \frac{1}{2}\frac{\Gamma}{v_{LC}} \phi_{LC} \varphi_{LC} S^2 - \frac{\phi_{LC}}{v_{LC}} \ln(Z) \qquad (1)$$

where the partition function $Z$ is given by:

$$Z = \int_0^1 \exp\left[\Gamma \phi S\left(x^2 - \frac{1}{3}\right)\right] \mathrm{d}x,$$

where *"P"* and *"LC"* stand for particle and liquid crystal respectively, $S$ is the uniaxial nematic order parameter, $v$ and $a$ are volume and surface-to-volume ratio (non-dimensionalized with respect to $v_{ref}$ and $v_{ref}^{2/3}$), $\phi$ and $\varphi$ are the volume and area fraction, $\chi$ is the mixing interaction parameter, $w$ is a binary nematic interaction parameter, $\Gamma$ is the Maier-Saupe quadrupolar interaction parameter, $R$ is the gas constant and $T$ is the temperature. $Z$ was accurately approximated by a polynomial expression in terms of $S$, and the coefficients of the expansion were obtained from a least-squares fitting of the numerical solution of the integral, as described in ref [25]. The following representative parameters were used to calculate the phase diagram: $v_c$ = 3, $v_p$ = 268, $a_{LC}$ = 4.66 , $a_P$ = 0.75, $\chi$ = 0.833/$T$, $\Gamma$ = 4.54/$T$, $w$ = -0.06667. These values were chosen such that the phase diagram shows a metastable L-L demixing, and the concentration in the equilibrium nematic phase is not too close to 1 (to avoid numerical problems). The temperature is normalized with respect to the nematic-isotropic transition temperature of the pure liquid crystal, $T_{NI}$.

Figure 3 shows the thermodynamic phase diagram used in this study, computed using Eq. (1). The construction of this phase diagram follows standard procedures: the equilibrium condition at each temperature is given by the equality of chemical potentials of each component and the minimization of the free energy with respect to the order parameter, in each phase. A more detailed analysis of related phase diagrams can be found in refs [23, 24]. Fig. 3 shows that, based on Eq. (1), two types of phases exist: an isotropic (I) phase , characterized by a zero scalar order parameter $S = 0$, and an uniaxial nematic (N) phase, with $S > 0$. Fig. 3 shows that an I + I coexistence region exists within the N+I coexistence region, and thus it describes a metastable equilibrium (this is denoted by $I^*+I^*$ in Fig. 3, the superscript stars are used to distinguish between the stable and metastable isotropic phases). This means that the free energy is minimal when the I and N phases coexist, but it has a local minimum for I + I coexistence. The phase separation spinodal (dotted line) is also shown in Fig. 3; inside this region, an isotropic phase is unstable with respect to composition fluctuations. The maximum temperature at which $I^* + I^*$ coexistence is possible is called the critical co-dissolution temperature ($T_c$). Above this temperature the two components are completely miscible in the isotropic phase. Below we consider phase transitions above and below $T_c$, as indicated by the points a-d in Fig. 3.

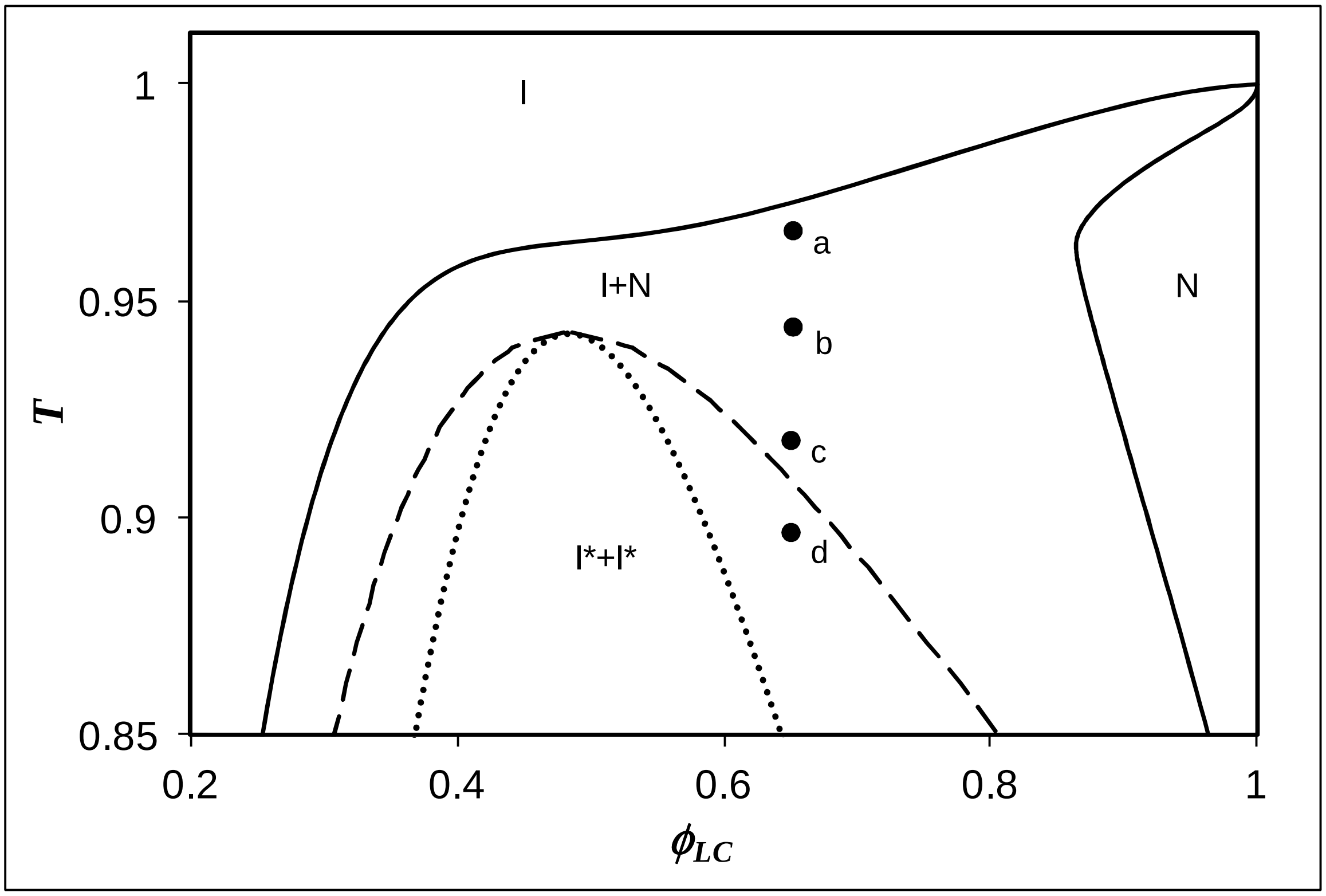

**Fig. 3.** Computed phase diagram based on Eq. (1); for: $v_c$=3, $v_p$ =268, $a_{LC}$=4.667, $a_P$=0.75 $\chi$ = 0.833/$T$, $\Gamma$= 4.54/$T$, $w$ = -0.06667. I and N represent the stable phases, I + N indicates the stable equilibrium and $I^*+I^*$ the metastable phase coexistence. Full lines indicate equilibrium binodal lines, dashed lines indicate the metastable isotropic-isotropic binodal. The dotted line indicates the phase separation spinodal. The positions a, b, c and d indicate the initial conditions used in the simulations.

The key issue to be established in this work is how the presence of this metastable state influences the phase transformation of an initially isotropic phase into demixed isotropic and nematic phases. Previous works have described the spontaneous formation of double interfaces, due to the appearance of metastable states, through a kinetic interface splitting mechanism [21,22]. The reader is referred to these previous works for technical details on the interfacial structure and kinematics. When the initial concentration of mesogen in the bulk I phase is smaller than the I-I critical concentration and the initial condition is in the $I^*$-$I^*$ coexistence region of the phase diagram, an $I^*$-$I^*$ interface (separating the two metastable phases) detaches from the initial I-N interface when the diffusion is faster than the ordering. This is ascribed to the fact that the N-I interface is controlled by the slower process (phase ordering), and consequently the metastable $I^*$-$I^*$ interface (purely diffusional), is faster. In the present case, where the initial condition is that the bulk concentration of mesogen is higher than the critical concentration, double interfaces arise but through a different mechanism, as will be shown in the following sections.

## 3. Dynamic model of phase transitions.

The dynamics of phase transitions in binary mixtures can be simulated considering the dynamics of a non-conserved order parameter coupled with a conserved variable (also known as model C) [21, 22]. In this work, time-dependent simulations in one-dimension are performed. The free energy density of a non-uniform system, $f$, is given by:

$$f = f_b + f_g \tag{2}$$

where the bulk contribution $f_b$ was given in the previous section, and the gradient contribution, $f_g$, is given by:

$$f_g = l_\phi \left(\frac{\partial \phi}{\partial x}\right)^2 + l_S \left(\frac{\partial S}{\partial x}\right)^2 + l_{S\phi} \frac{\partial \phi}{\partial x}\frac{\partial S}{\partial x} \tag{3}$$

where $l_\phi$, $l_S$, and $l_{S\phi}$ are standard gradient energy coefficients [21; 22]. The dynamic equations obtained from Eqns (1-3) in one dimension are:

$$\frac{\partial S}{\partial t} = M_S \left( -\frac{\partial f_b}{\partial S} + \frac{\partial}{\partial x} \frac{\partial f_g}{\partial \left(\frac{\partial S}{\partial x}\right)} \right) \tag{4}$$

$$\frac{\partial \phi}{\partial t} = \frac{\partial}{\partial x}\left( M_{\phi} \frac{\partial \mu}{\partial x} \right) = M_{\phi} \frac{\partial^2}{\partial x^2}\left( \frac{\partial f_b}{\partial \phi} - \frac{\partial}{\partial x} \frac{\partial f_g}{\partial \left( \frac{\partial \phi}{\partial x} \right)} \right) \tag{5}$$

where $M_S$ and $M_{\phi}$, the mobilities for phase ordering and mass diffusion, are assumed to be constant. As usual, the chemical potential, $\mu$, is given by the functional derivative of the free energy with respect to composition and the generated mass flux j is given by $j = M_{\phi}\, \partial\mu/\partial x$. Time and position are expressed in units of the characteristic times and length scales, $\tau = v_{ref}/(M_S R T_{NI})$ and $l = [v_{ref} l_{\phi}/(RT_{NI})]^{1/2}$. Comsol Multiphysics was used to solve Eqns. (4) and (5), with quadratic Lagrange basis functions; standard numerical techniques were used to ensure convergence and stability. Transient simulations were performed in one dimension, using soft step functions in concentration and order parameter as initial conditions, where the concentration of the bulk isotropic phase and temperature are given by the points a-d in Fig. 3, and the initial composition and order parameter in the nematic phase correspond to the equilibrium nematic phase at each selected temperature.

Next we state a number of facts on diffusion-control that are used in the analysis of numerical results. In a coupled variable system as found in this work, when one variable is much slower than the other one, the phase transition will be kinetically controlled by this slow variable. It is well known that COP and NCOP follow different dynamic growth laws. In the case of NCOP [Eq. (3)], the velocity of the interface is constant [16, 17], while in the case of COP [eqn. (4)], the velocity is proportional to $t^{-1/2}$ [19, 20]. Hence for the COP and NCOP binary mixture considered in this paper, diffusion-control is expected at sufficiently long times. Diffusion-controlled processes are often described by Fick's law. In its traditional formulation the flux is expressed as $j = -D\, \partial\phi/\partial x$, and if the diffusivity $D$ is constant, the evolution equation is:

$$\frac{\partial \phi}{\partial t} = D \frac{\partial^2 \phi}{\partial x^2} \tag{6}$$

Eqn. (6) can be considered a sharp-interface equivalent to Cahn-Hilliard model [Eqn. (5)], in that it neglects the gradient contributions (thus is applicable only in bulk phases), and with $D = M_{\phi} \partial^2 f_b / \partial \phi^2$. If the mobility is constant, in general the diffusivity will not be constant as the value of $\partial^2 f_b / \partial \phi^2$ usually depends on composition. Nevertheless, the assumption of constant diffusivity works well when $\partial^2 f_b / \partial \phi^2$ is not a strong function of composition. When the diffusivity is constant, Eqn. (6) yields analytical solutions in the several relevant cases used below.

## 4. Phase transitions controlled by ordering or with mixed dynamics.

In previous work we have analyzed the general features of the case of mixed dynamics and phase-ordering control [21, 22]. The velocity of the interface decreases with an exponent smaller than the diffusive value of 1/2**,** and we argued that this value would always be approached asymptotically. In this case, as the interface is not at equilibrium, the concentrations of each phase at the interface do not correspond to the equilibrium concentrations. When the process is controlled by ordering, diffusion is infinitely fast, the concentration in the isotropic phase is uniform, and the interfacial concentration is the same as in the bulk. Because of the relative location of the initial conditions with respect to the phase diagram, the interface splitting kinetic process observed in previous work does not take place in this case.

For higher relative mobility or for longer times, the phase transition will show mixed dynamics and it will evolve towards a regime of diffusional control. The composition at the interface in the isotropic phase decreases with time from the initial bulk value to the equilibrium value. When the temperature is above $T_c$, this decrease is continuous and smooth, but the situation is different below this temperature.

Fig. 4a shows the concentration profiles for case c, shown in Fig. 3, with $M_\mathrm{r} = M_\phi/(M_\mathrm{Q} l^2) = 10^5$, for different times. Initially the decrease of the concentration at the interface is slow and smooth, but as soon as this concentration hits the I-I spinodal line, it decreases abruptly and a double front is formed (see lowest “well-like” profile in Fig. 4a). This can be seen more clearly in Fig. 4(b) where the concentration at the interface is shown as a function of time; note the drop around $t$ = 200. This process does not have a kinetic origin; it is essentially a phase separation in the diffusion depletion layer, and this mechanism has several differences compared to the kinetic interface splitting described in previous works, as discussed in the next section.

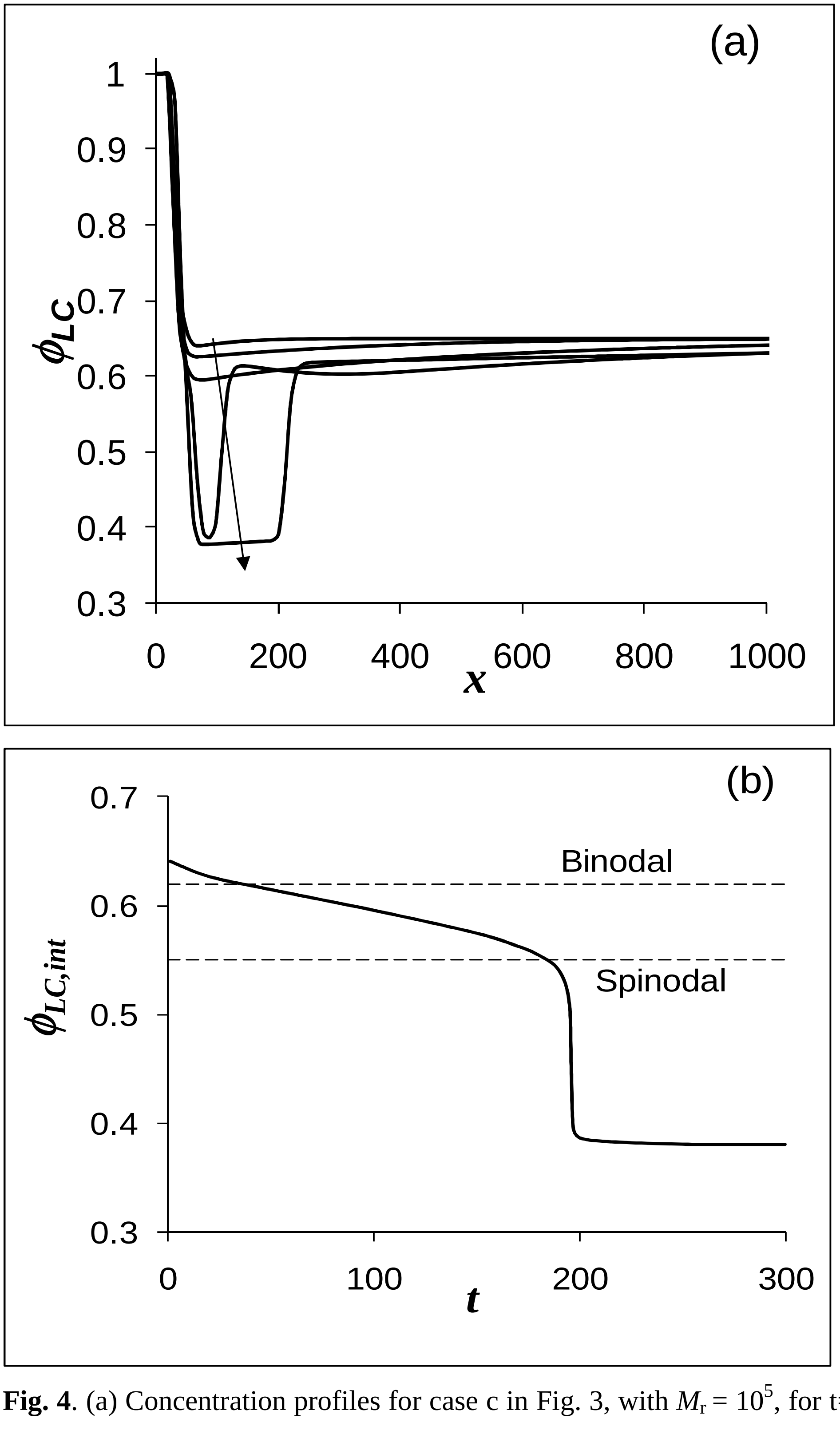

**Fig. 4**. (a) Concentration profiles for case c in Fig. 3, with $M_r = 10^5$, for t= 1, 20, 100, 190, 200, and 300, increasing in the direction of the arrow. (b) Concentration in the isotropic phase at the interface as a function of time. The binodal and spinodal concentration are indicated.

## 5. Dynamics of phase transitions controlled by mass diffusion

### 5.1 – Model C Results

Fig. 5(a)-(d) show representative concentration $\phi$ and order parameter $S$ profiles for the cases a-d shown in Fig. 3, for $M_r$ = 1000. The insets in this figure show the location of characteristic interface compositions in the phase diagram. For high temperature [Fig. 5(a)], the typical diffusion case is observed, where the concentration in the depletion layer increases monotonically with a smoothly decreasing curvature to the bulk value. This profile and the evolution of the interface can be well represented by Fick´s law (Eq.(6)), with a constant diffusivity.

When the temperature approaches $T_c$ from above, an inflection point is developed in the depletion layer, as shown in Fig. 5(b). This fact can be understood if the stationary solution of the diffusion problem is considered. Although the stationary solution is never reached, the system slowly evolves towards it, so extrapolating the stationary solution to the dynamic regime is reasonable, at least qualitatively. The stationary solution of the evolution equation [Eqn. (5)] corresponds to a uniform mass flux. As stated in section 3, the mass flux is given by $j = M_\phi d\mu/dx$, which, neglecting gradient energy contributions, reduces to $j = M_\phi \left(\partial^2 f_b / \partial\phi^2\right)\left(\partial\phi / \partial x\right)$. The stationary solution corresponds then to a linear profile of chemical potential, or equivalently a constant value of the product $\left(\partial^2 f_b / \partial\phi^2\right)\left(\partial\phi / \partial x\right)$. The shape of the stationary concentration profile is then directly related to the shape of the curve $\mu$ as a function of $\phi$, so an inflection point in the stationary concentration profile appears when $\mu(\phi)$ has an inflection point. The formation of a marked inflection in $\mu(\phi)$ is associated with strong positive deviations from the ideal solution (positive excess free energy) produced by a large positive interaction terms in the free energy density (Eq.(1)). Extreme positive deviations produce phase separation, at which $\left(\partial^2 f_b / \partial\phi^2\right)$ becomes zero at the critical co-dissolution point (the inflection point in $\mu(\phi)$ is horizontal). At this point, if the gradient energy contribution is zero, the slope of the inflection point in the stationary concentration profile would be infinite.

Fig. 6 shows the bulk isotropic contribution to the chemical potential as a function of concentration, for the temperatures corresponding to each case. The range of concentration comprised by the depletion layer is shown as a solid line. It can be seen that, for case a, even when an visible inflection point exist in the curve, it is outside the relevant range of concentrations; in this range the chemical potential is almost a linear function of the composition so the typical profile is observed. In case b, a clear inflection point with a slope close to zero is comprised within the range of concentrations of the depletion layer, which leads to the inflection point in the concentration profile shown in Fig. 5(b).

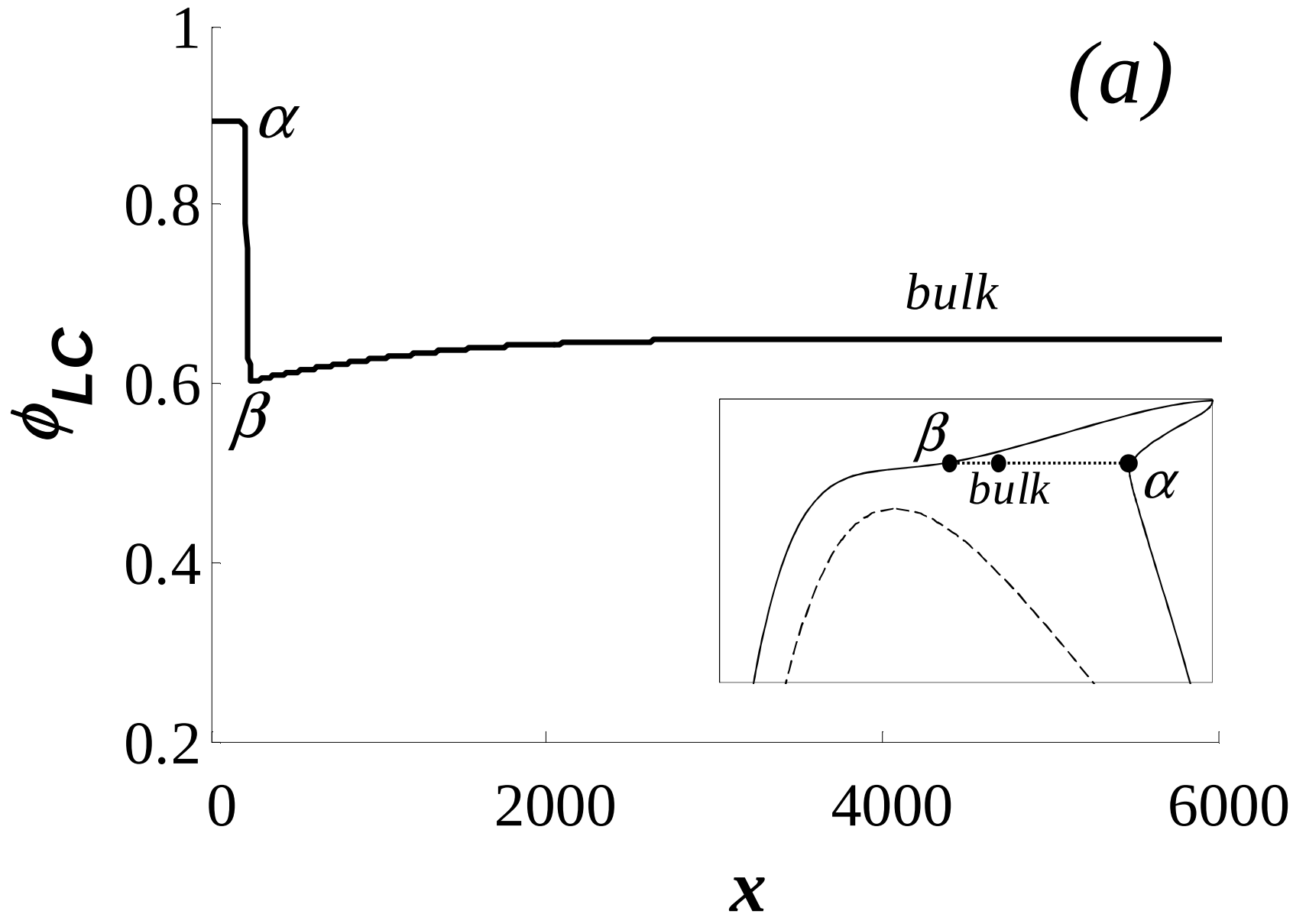

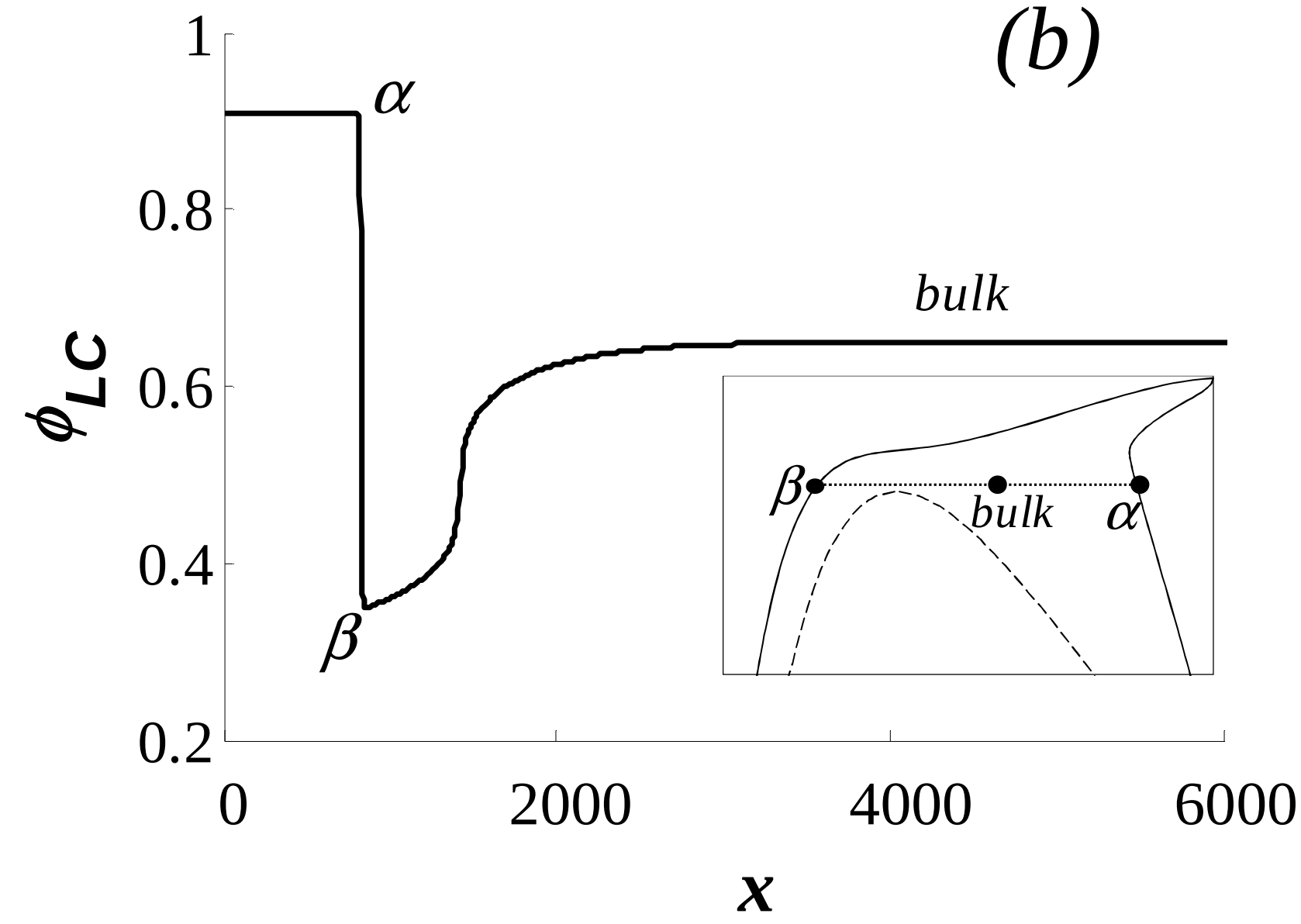

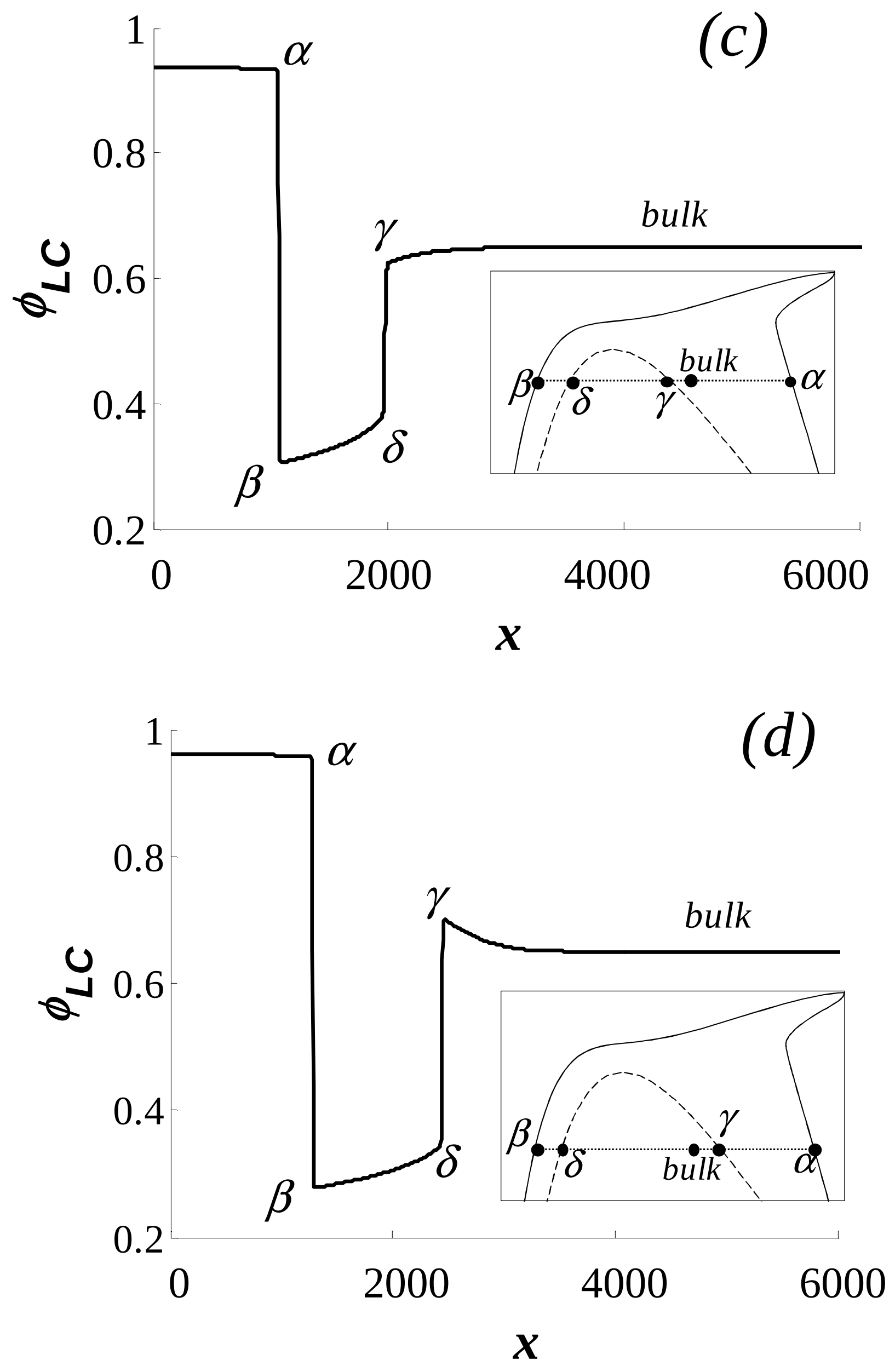

**Fig. 5.** Concentration and scalar order parameter profiles, for $M_r$ = 1000, for cases a, b, c and d defined in Fig. 3, at $t$ = $10^4$. The insets in each figure show the situation of the characteristic points $(\alpha, \beta, \delta, \gamma, bulk)$ in the phase diagram.

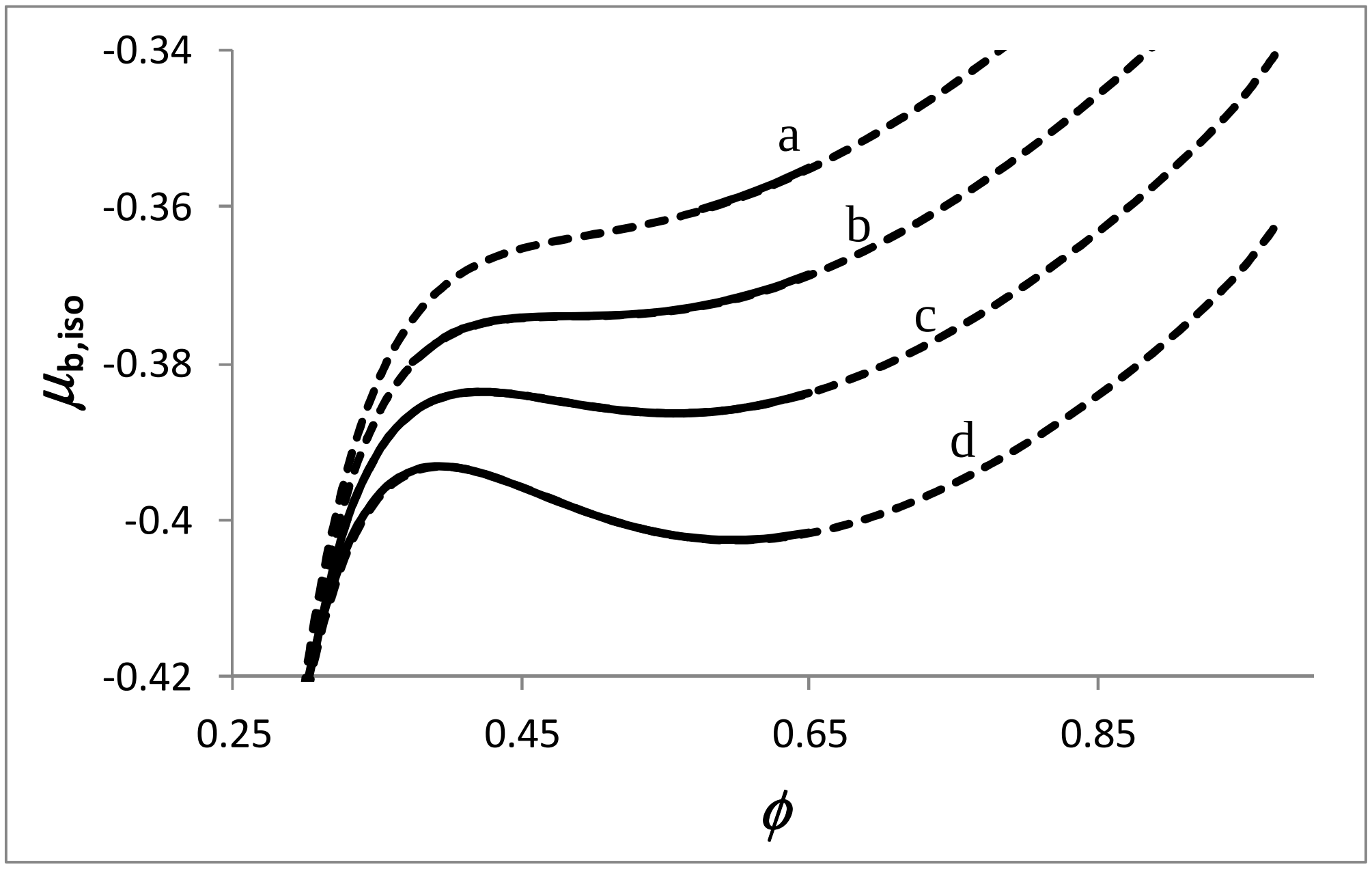

**Fig. 6.** Bulk isotropic contribution to the chemical potential as a function of composition, for the temperatures corresponding to each of the cases a, b, c and d, as indicated. The dashed lines represent the curve in the entire concentration range, while the full lines show the concentration range comprised in the depletion layer (between the concentrations at the bulk and at the interface)

When the temperature is below the critical temperature, as discussed in the previous section, the concentration in the depletion layer crosses the I-I coexistence region. The curve $\mu(\phi)$, as seen in Fig. 6, has not only an inflection point but also a local minimum and a local maximum, with $\left(\partial^2 f_b / \partial\phi^2\right)$ becoming negative in some concentration range, which means thermodynamic instability and phase separation. This leads to a phase separation in the depletion layer, which develops as soon as the concentration in the depletion layer hits the unstable (spinodal) region, as shown in the previous section. The corresponding concentration profiles are shown in Fig. 5(c) and (d). The difference between cases c and d is that the temperature in case c is above the binodal and in case d is below or, equivalently, the concentration in the bulk is outside the binodal in case c and inside in case d. In case c, the mass flux has a uniform direction, from the isotropic bulk to the nematic phase, while in case d, there is a flux from the second interface to the bulk. Note that the velocity of this interface is determined by the difference between the fluxes at the right and left of the interface, so a change in the sign of one flux does not imply a change in the direction of the interface motion.

This double-front formation mechanism is different to that observed previously [21, 22] for a concentration lower than the critical point. Since the mass diffusion is faster than ordering kinetics in this previous case, the interface splits into a fast diffusive interface and a slow interface kinetically controlled by ordering. The metastable phase was formed through this splitting mechanism in between the two interfaces, and an isotropic phase with the equilibrium concentration was not present at this stage. This split state had a finite lifetime, so eventually the two interfaces merged again, the intermediate metastable phase disappeared, and the system kept evolving to equilibrium. In the present case, the second interface is formed by a thermodynamic effect. The second front appears when the concentration at the interface hits the spinodal region, due to a phase separation in the depletion layer rather than a kinetic interface splitting. The stable phase, and not the metastable one, is formed between the two interfaces, so the split state will not have a finite lifetime. Instead, the second interface will be "pushed away" towards infinity as the stable phases grow. Another important difference between the present and previous case [21, 22] is that the double front appears even if the initial conditions of ($\phi$, $T$) are not inside the binodal (as in case c), the condition for this to happen is that $T < T_c$ (unlike the previous case, were the system has to be inside the binodal for the splitting to occur). Finally, in the present case the double front appears when the dynamics is controlled by diffusion, while the kinetic mechanism of interface splitting studied previously [21, 22] was observed when the dynamics are controlled by ordering.

The case analyzed by Evans et al [19 20], which is for pure conserved parameters (and thus is a diffusive process like the present case), is analogous to the "kinetic interface splitting" and also has some important differences with the present results. In their case, the interface splitting was not spontaneous (it required a finite perturbation), and the separation between the interfaces could increase monotonically or the interfaces could merge again after a finite time, depending on the degree of undercooling. In the present case, the formation of the second interface is spontaneous (as shown in the previous section), and, as will be proven in the next section, the separation between the interfaces increases monotonically.

### 5.2. Fick´s sharp-interface analytical diffusion model.

As discussed before, the Cahn-Hilliard model (Eqn. 4) can be related to the classical Fick's diffusion model by setting $D = M_\phi\left(\partial^2 f_b / \partial\phi^2\right)$, and assuming that the gradient energy contribution to the mass flux is negligible. This is a good assumption as in general the gradient contribution becomes appreciable only across the interfaces. In order to find analytical solutions, the diffusivity is assumed to be constant. As shown above, the split interface is ascribed to very strong deviations from ideal behavior (strong variations and sign change of $\partial^2 f_b / \partial\phi^2$). Nevertheless, in the sharp-interface formulation, the evolution equation [Eqn. (6)] can be solved analytically for concentrations outside the $I^*$-$I^*$ region of the phase diagram, where the deviations are small or moderate, so this assumption can still be made.

The system is divided in three regions (I-II-III), as shown schematically in Fig. 7: (I) corresponding to the nematic phase, (II) corresponding to the "intermediate" (stable) isotropic phase, and (III) corresponding to the "bulk" isotropic phase. In region I we assume that there is no flux, as the concentration profile is rather flat, and in regions II and III, Fick´s diffusion equation with constant diffusivity can be applied. Similar formulations for double-fronts in diffusion problems can be found in Evans et al [19, 20], for the case when the intermediate phase is the metastable one, and in Peng et al [26], where they present a detailed mathematical analysis for a case of a double front in a sublimation – desorption problem.

The system of equations describing the evolution of the system, with the above-stated conditions, is:

$$\frac{\partial \phi^{II}}{\partial t} = D^{II} \frac{\partial^2 \phi^{II}}{\partial x^2} ; \qquad \frac{\partial \phi^{III}}{\partial t} = D^{III} \frac{\partial^2 \phi^{III}}{\partial x^2} \tag{7}$$

and the boundary and initial conditions and mass balances at the interfaces are:

$$\begin{gathered} \phi^{II}\Big|_{x=X_1} = \phi_\beta ; \quad \phi^{II}\Big|_{x=X_2} = \phi_\delta ; \quad \phi^{III}\Big|_{x=X_2} = \phi_\gamma ; \\ \phi^{III}\Big|_{x=\infty} = \phi_{bulk} ; \qquad \phi^{III}\Big|_{t=0} = \phi_{bulk} \; \phi^{III}\Big|_{t=0} = \phi_{bulk} \\ -\left( D^{II} \frac{d\phi^{II}}{dx}\bigg|_{x=X_2} - D^{III} \frac{d\phi^{III}}{dx}\bigg|_{x=X_2} \right) = \left(\phi_\delta - \phi_\gamma\right) \frac{dX_2}{dt} ; \\ D^{II} \frac{d\phi^{II}}{dx}\bigg|_{x=X_1} = \left(\phi_\alpha - \phi_\beta\right) \frac{dX_1}{dt} \end{gathered} \tag{8}$$

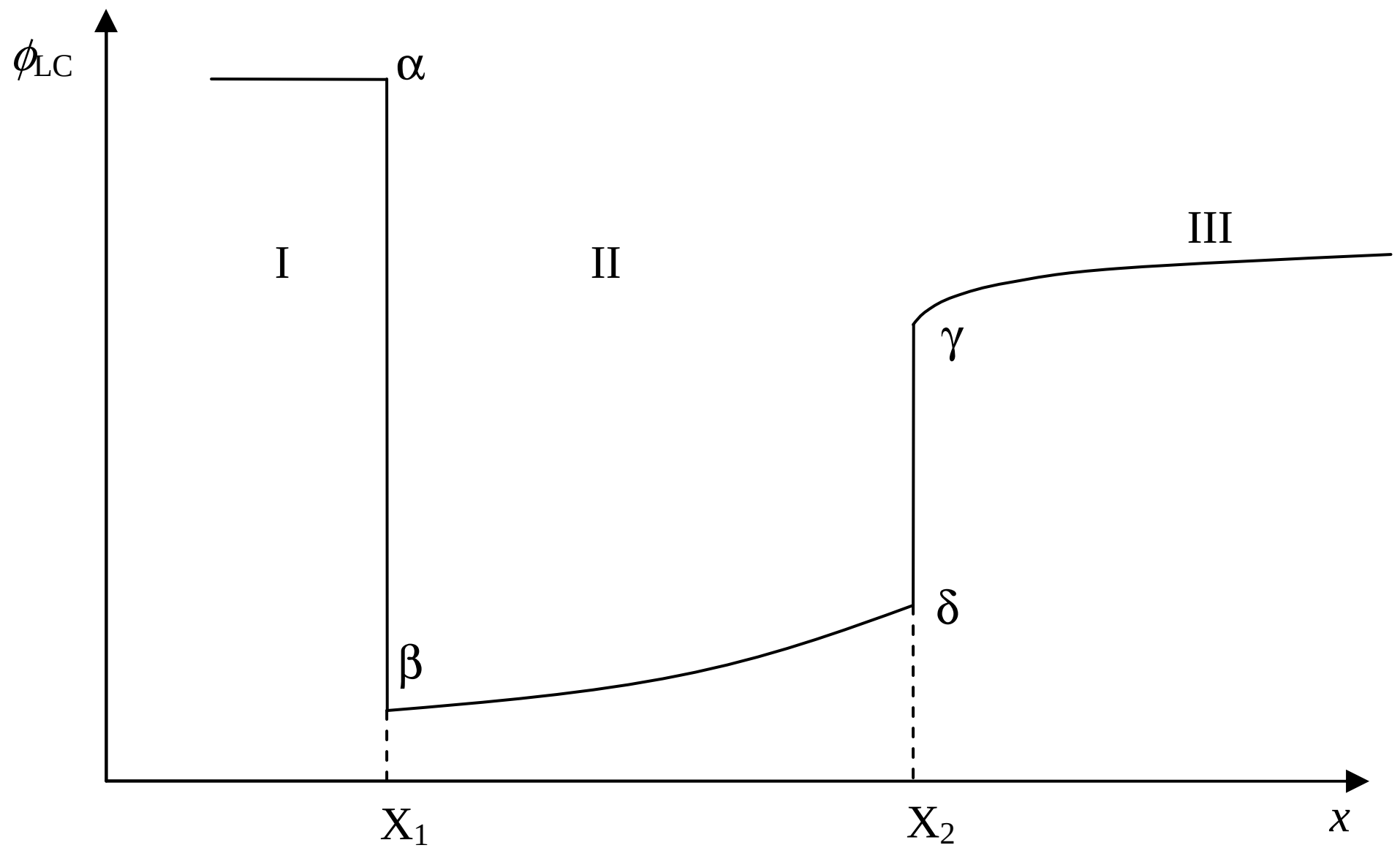

**Fig. 7.** Schematics of concentration profiles in a diffusional transformation with a double front (located at $X_1$ and $X_2$), corresponding to case c; see Fig. 5(c) for an actual computation. The characteristic concentrations at the interfaces are labeled with Greek letters.

Initially at $t$=0, $X_1$=$X_2$=0 and $\phi^{II}$ is undefined. Note that we assume constant diffusivity within each region, but each region can have a different diffusivity. Also, as the system is controlled by diffusion, the interfaces are in equilibrium. The above-stated model (Eq. (7, 8)) can be solved by performing the transformation $\eta = 0.5x(Dt)^{-1/2}$, and assuming $X_1 = 2A(Dt)^{1/2}$, $X_2 = 2B(Dt)^{1/2}$, where $D$ has the dimensions of a diffusivity. If this is replaced in Eqns. (7) and (8) (See Peng et al. [26] for details), the following solution is obtained:

$$\phi^{II} = \frac{\phi_\delta - \phi_\beta}{erf\left(B\sqrt{\frac{D}{D^{II}}}\right) - erf\left(A\sqrt{\frac{D}{D^{II}}}\right)}\left[erf\left(\eta\sqrt{\frac{D}{D^{II}}}\right) - erf\left(A\sqrt{\frac{D}{D^{II}}}\right)\right] + \phi_\beta \tag{9a}$$

$$\phi^{III} = \frac{\phi_{bulk} - \phi_\gamma}{1 - erf\left(B\sqrt{\frac{D}{D^{III}}}\right)}\left[erf\left(\eta\sqrt{\frac{D}{D^{III}}}\right) - erf\left(B\sqrt{\frac{D}{D^{III}}}\right)\right] + \phi_\gamma \tag{9b}$$

$$\left(\phi_\delta - \phi_\gamma\right)B = \frac{1}{\sqrt{\pi}}\left[\frac{\phi_{bulk} - \phi_\gamma}{1 - erf\left(B\sqrt{\frac{D}{D^{III}}}\right)}\sqrt{\frac{D}{D^{III}}}\exp\left(-\frac{D}{D^{III}}B^2\right) - \frac{\phi_\delta - \phi_\beta}{erf\left(B\sqrt{\frac{D}{D^{II}}}\right) - erf\left(A\sqrt{\frac{D}{D^{II}}}\right)}\sqrt{\frac{D}{D^{II}}}\exp\left(-\frac{D}{D^{II}}B^2\right)\right] \tag{10a}$$

$$\left(\phi_\alpha - \phi_\beta\right)A\sqrt{\frac{D}{D^{II}}} = \frac{\phi_\delta - \phi_\beta}{erf\left(B\sqrt{\frac{D}{D^{II}}}\right) - erf\left(A\sqrt{\frac{D}{D^{II}}}\right)}\frac{1}{\sqrt{\pi}}\exp\left(-A^2\frac{D}{D^{II}}\right) \tag{10b}$$

Note that if $D = RT_{NI}/v_{\mathrm{ref}}M_{\phi}$ is defined, $D^{\mathrm{II}}/D$ and $D^{\mathrm{III}}/D$ are just the representative values of $\frac{\partial^2}{\partial\phi^2}\left(\frac{v_{ref}}{RT}f_b\right)$ in each region.

Some important conclusions can be extracted from this analytical model that cannot be explicitly seen in the continuous interface model solved numerically. It can be seen from Eqn. (10b), that for A to be positive, B > A holds and, as $X_2 - X_1 = 2(B\text{-}A)(Dt)^{1/2}$, the separation between the interfaces increases monotonically and unconstrained with time. Also, Eqn. (10a) explicitly shows that, depending on whether the initial condition is outside the binodal ($\phi_\gamma > \phi_{bulk}$), or inside ($\phi_\gamma < \phi_{bulk}$), the mass flux in region III at the interface will be negative or positive, contributing to accelerate or decelerate the interface.

As a quantitative comparison between the two models, Fig. 8 shows the separation between the two interfaces, $X_2 - X_1$, as a function of time, for cases c and d and using different values of relative mobility, calculated from simulations [Eqns. (4) and (5)] and from the sharp-interface analytical model [Eqns.(7)-(10)]. The average value of $\partial^2\left(v_{ref}f_b/RT\right)/\partial\phi^2$ evaluated at the boundaries of each region was used. It can be seen that the assumption of constant diffusivity in each region gives a reasonable result.

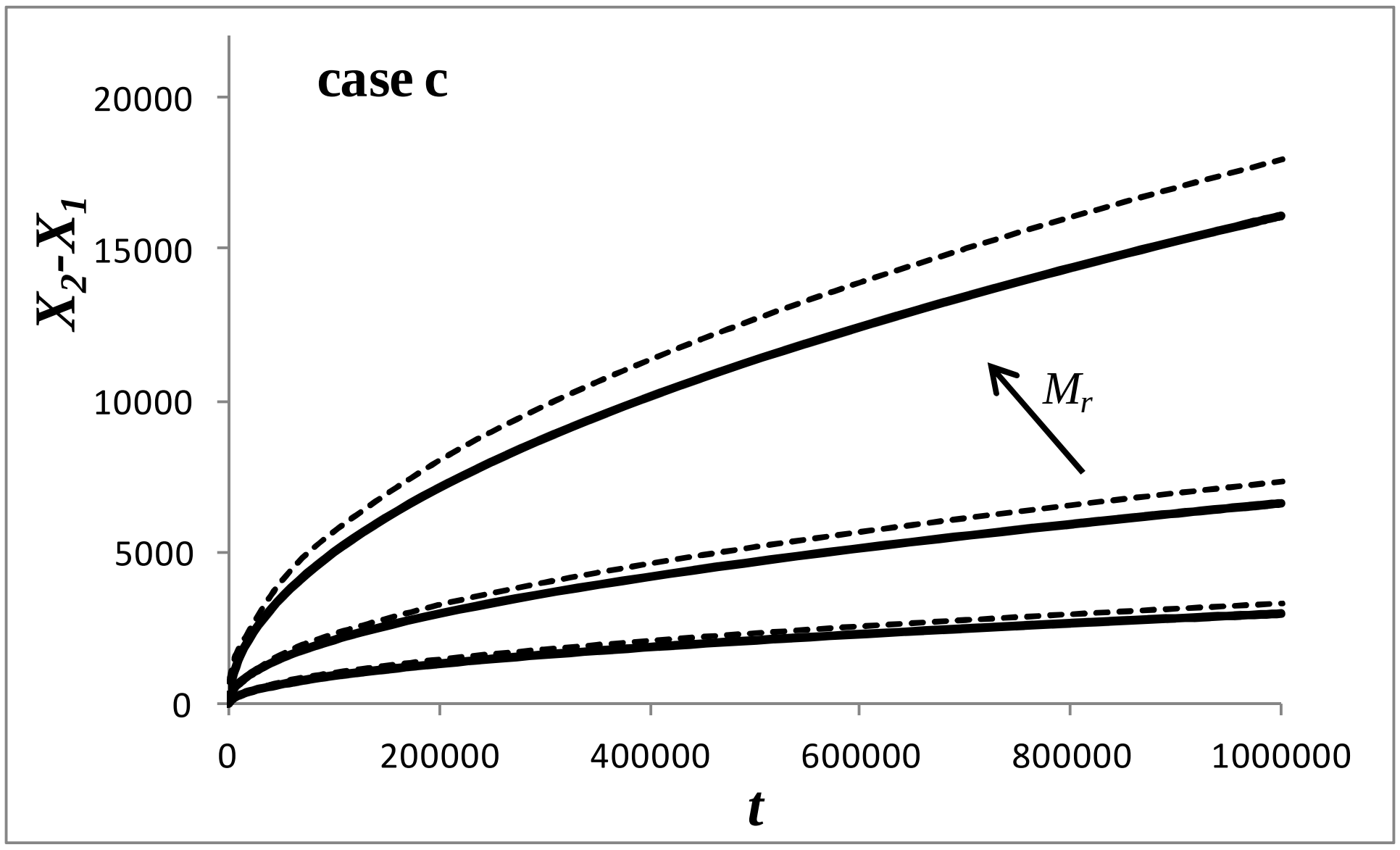

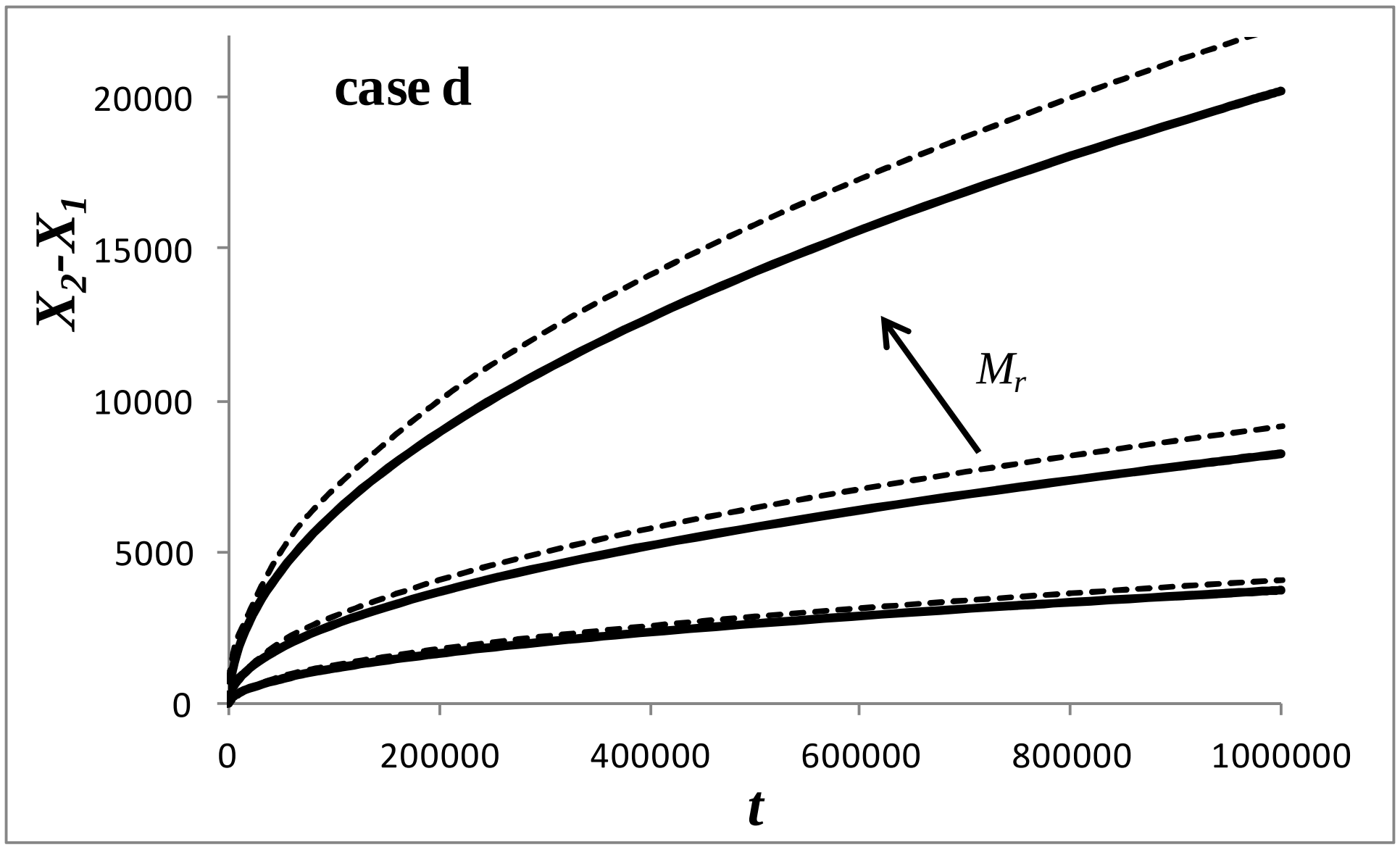

**Fig. 8.** Evolution of the separation between interfaces, for cases c and d, for $M_r$ = 100, 500 and 3000 (increasing in the direction of the arrow), from computational results from model C (full lines), and analytical results from the sharp-interface model (dashed lines).

## 6. Conclusions

This paper presented theory (Eqn. 1), analysis (Eqn. 9-10), and computation (Figs. 3-9) of 1D phase transition kinetics in a binary mesogen-nonmesogen mixture whose thermodynamic state is defined by one conserved (concentration) and one nonconserved nematic order parameter, and whose thermodynamic phase diagram (Fig. 3) is uniquely distinguished by the presence of a buried metastable equilibrium. The 1D dynamics of the isotropic-nematic phase transition in this binary system exhibiting simultaneous chemical demixing and nematic phase ordering was modeled, by means of numerical continuous interface (phase field) modeling, and sharp-interface Fick´s diffusion theory, in conditions where a metastable isotropic-isotropic phase equilibrium exists (Fig. 3). The growth of an initially existing nematic phase was analyzed, by following the dynamics of the nematic-isotropic interface, when the concentration of mesogen in the isotropic phase is higher than the isotropic-isotropic critical concentration. In the case of a diffusional phase transition, when the temperature approaches the isotropic-isotropic critical temperature from above, a strongly atypical profile was observed, with the formation of an inflection point in the depletion layer (Fig. 5), associated with positive deviations from ideal solution. For temperatures below the critical temperature, a phase separation within the diffusion depletion layer was observed, leading to the formation of a double front.

This novel thermodynamic mechanism is driven by the fact that the metastable isotropic-isotropic binodal is embedded within the concentrations of the bulk and at the interface. As the isotropic phase between the two interfaces is the equilibrium one, this phase grows with time, which means that the separation between the two interfaces grows unconstrained with time. This is in contrast to the kinetic mechanism described previously [21, 22], for concentrations in the isotropic phase lower than the critical one, where the interface splits in two, forming a metastable phase between the two interfaces. In that case [21, 22], the split state had a finite lifetime, and the condition for this to happen was that the temperature must be below the temperature of the binodal line at that concentration (inside the region of metastable equilibrium). Despite the fact that the overall concentration profile strongly departs from the classical case corresponding to Fickian diffusion with constant diffusivity, the process can still be described with this model, provided that it is applied to each phase separately and the equations are coupled at the interface (Figure 8).

The integration of computational modeling and transport theory provides a comprehensive description of the role of metastability when demixing in binary mesogen-nonmesogen solutions coexists with nematic phase ordering. It is shown that the growth of nematic domains under metastability effects will exhibit complex interfaces whose structure and spatial extension depends on thermodynamics and mobilities. As this process depends on some general thermodynamic and kinetic conditions (existences of a metastable L-L equilibrium, kinetics controlled by diffusion, concentration above the L-L critical point), it is possible in any material system given that those conditions are met. Nevertheless, it might be difficult to observe experimentally due to the competition with other mechanisms of formation of the metastable phase (spinodal decomposition, faster nucleation rate), and the presence of multiple, polydisperse domains with irregular shapes. This mechanism is expected to be more relevant (and easier to observe), for example, in directional growth experiments, where a single planar interface is generated. Even in other set-ups where direct observation of the metastable phase can be experimentally difficult, it can still have a significant impact on phase transition dynamics and on the structure of the material.

**Acknowledgements**

This work is supported by a grant from Le Fonds de Recherche du Québec – Nature et Technologies (FQNRT). ADR acknowledges partial support from the U.S. Office of Basic Energy Sciences, Department of Energy, grant DE-SC0001412. ERS acknowledges partial funding provided by the National Research Council of Argentina (CONICET), and University of Mar del Plata (UNMdP).